\documentclass[conference]{IEEEtran}

\IEEEoverridecommandlockouts
\usepackage{cite}
\usepackage{amsmath,amssymb,amsfonts}
\usepackage{algorithmic}
\usepackage{graphicx}
\usepackage{textcomp}
\usepackage{xcolor}
\usepackage{soul}
\def\BibTeX{{\rm B\kern-.05em{\sc i\kern-.025em b}\kern-.08em
    T\kern-.1667em\lower.7ex\hbox{E}\kern-.125emX}}

\begin{document}

\title{Guardians of the Quantum GAN\\

}

\author{\IEEEauthorblockN{Archisman Ghosh}
\IEEEauthorblockA{\textit{CSE Department} \\
\textit{Penn State University}\\
State College, PA, USA\\
apg6127@psu.edu}
\and
\IEEEauthorblockN{Debarshi Kundu}
\IEEEauthorblockA{\textit{CSE Department} \\
\textit{Penn State University}\\
State College, PA, USA \\
dqk5620@psu.edu}
\and
\IEEEauthorblockN{Avimita Chatterjee}
\IEEEauthorblockA{\textit{CSE Department} \\
\textit{Penn State University}\\
State College, PA, USA \\
amc8313@psu.edu}
\and
\IEEEauthorblockN{Swaroop Ghosh}
\IEEEauthorblockA{\textit{School of EECS} \\
\textit{Penn State University}\\
State College, PA, USA \\
szg212@psu.edu}}

\maketitle

\begin{abstract}
Quantum Generative Adversarial Networks (qGANs) are at the forefront of image-generating quantum machine learning models. To accommodate the growing demand for Noisy Intermediate-Scale Quantum (NISQ) devices to train and infer quantum machine learning models, the number of third-party vendors offering quantum hardware as a service is expected to rise. This expansion introduces the risk of untrusted vendors potentially stealing proprietary information from the quantum machine learning models. To address this concern we propose a novel watermarking technique that exploits the noise signature embedded during the training phase of qGANs as a non-invasive watermark. The watermark is identifiable in the images generated by the qGAN allowing us to trace the specific quantum hardware used during training hence providing strong proof of ownership. To further enhance the security robustness, we propose the training of qGANs on a sequence of multiple quantum hardware, embedding a complex watermark comprising the noise signatures of all the training hardware that is difficult for adversaries to replicate. We also develop a machine learning classifier to extract this watermark robustly, thereby identifying the training hardware (or the suite of hardware) from the images generated by the qGAN validating the authenticity of the model. We note that the watermark signature is robust against inferencing on hardware different than the hardware that was used for training. We obtain watermark extraction accuracy of $100\%$ and $\sim90\%$ for training the qGAN on individual and multiple quantum hardware setups (and inferencing on different hardware), respectively. Since parameter evolution during training is strongly modulated by quantum noise, the proposed watermark can be extended to other quantum machine learning models as well.

\end{abstract}


\section{Introduction}
Generative Adversarial Networks (GANs) are a pivotal neural network architecture used in computer vision \cite{gan1}. A classical GAN uses two neural networks, a generator and a discriminator. The role of the generator is to create synthetic data that mimics real data from the training dataset, while the discriminator evaluates this data against the training dataset to determine its authenticity. This method of training helps generate new data with the same statistics as the training dataset. Classical GANs find numerous multiple real-world applications like text-to-image generation \cite{hu2022text}. They also help in other tasks like increasing image resolution \cite{ledig2017photorealistic}, and image inpainting \cite{wang2022highfidelity} to name a few. 
With the increasing complexity of GANs, research has shifted towards optimizing their training and design with the help of quantum computers. Recent research indicates \cite{Dallaire_Demers_2018} that quantum GANs (qGANs) may have exponential advantages over classical GANs in optimizing the number of parameters in the generator and discriminator models. 

To cope with the high demand for designing and training quantum machine learning (QML) models quantum hardware providers have taken the same route as the classical machine learning engineers \cite{sharma2022analysis}, i.e., providing quantum machine learning hardware as a service to quantum circuit designers for training and inferencing complex QML models like qGANs. In the near future, the number of third-party vendors providing quantum hardware as a service is bound to increase which will reduce the overall cost of training QML models. 
Many third-party quantum cloud service providers are evolving around the globe some of which are from less trustworthy countries\cite{upadhyay2023trustworthy}. Such quantum service providers may offer computing at a cheaper rate in addition to readily available hardware. Both of these factors could be motivating for users who would like to obtain results quickly at a cheaper cost to boost their profit margins.  
However, owing to the importance of high-quality training datasets and the complex circuit design of the qGAN models, untrusted third-party vendors or rogue adversaries will be incentivized to steal trained qGAN models, leaking sensitive data from the training dataset or use the stolen models in an unauthorized way.

\subsection{Motivation}
\begin{table}
\centering
\caption{Pricing comparison for execution on real quantum hardware}
\begin{tabular}{c|c|c}

\textbf{Hardware Provider} & \textbf{QPU} & \textbf{Rate(per hr)}  \\ \hline \hline
IonQ & Aria & \$7000 \\ \hline
QuEra & Aquila & \$2500  \\ \hline
Rigetti & Aspen-M-3 & \$3000 \\ \hline
IBM & Eagle & \$5760 (\$1.6/s) \\ \hline \hline

\end{tabular}

\label{tab:hardware_cost}
\end{table}

A trained qGAN circuit possesses many intellectual properties (IP) such as the quantum circuit model architecture, the trained model, and weights, since developing a quantum circuit design and training it on quantum hardware requires considerable time and resources (Table \ref{tab:hardware_cost}).  


During the inferencing phase, if this trained qGAN model is sent to an untrusted quantum cloud provider—or even falls into the hands of a rogue adversary within a trusted provider—it risks being stolen \cite{szyller2023good}. Counterfeit qGAN models, being inexpensive to deploy and operate, can yield significant profits, thereby enticing untrusted vendors and rogue adversaries to engage in IP theft. The motivation to steal these models is strong because preparing a training dataset is costly, and the training process on quantum hardware is both time-intensive and expensive. Without effective security measures, the incidence of counterfeit models could surge, making it challenging to distinguish between genuine and counterfeit models and potentially destabilizing the quantum service market.  This issue is analogous to the counterfeiting of hardware chips by untrusted fabrication houses and the theft of classical machine learning models by untrusted classical cloud providers. Various strategies have been developed to combat model theft to enhance the security of the classical GANs by embedding external signatures in the generated images to support steganography \cite{Yuan2022}. However, these traditional defense methods involving image steganography may prove ineffective for images produced by quantum adversarial networks. The intrinsic noise in quantum hardware could result in data loss while embedding the signature, thereby risking its integrity and effectiveness.

\begin{figure}
    \centering
    \includegraphics[width=1\linewidth]{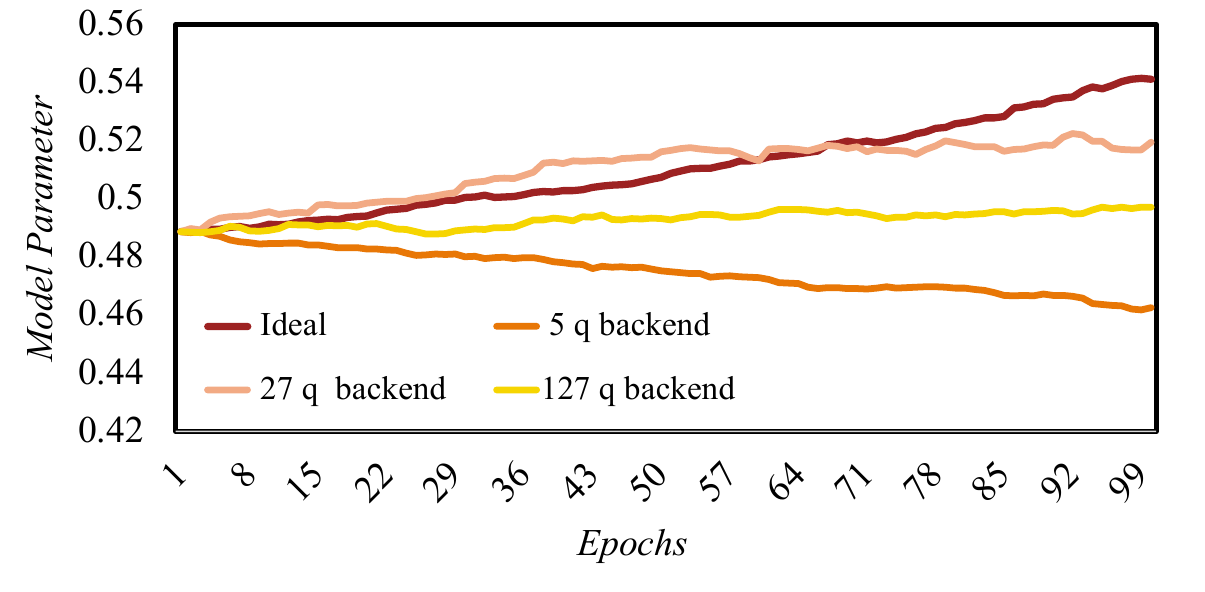}
    \caption{The plot describes the fluctuation in the training of the same parameter in the quantum generator of the qGAN while training on ideal hardware (noiseless), and different noisy quantum hardware - IBM Athens(5q), IBM Jakarta(7q), IBM Kolkata(27q), IBM Washington(127q).
     } 
    \label{fig:param}
    \vspace{-10pt}
\end{figure}

\subsection{Contribution}
\begin{figure}
    \centering
    \includegraphics[width=1\linewidth]{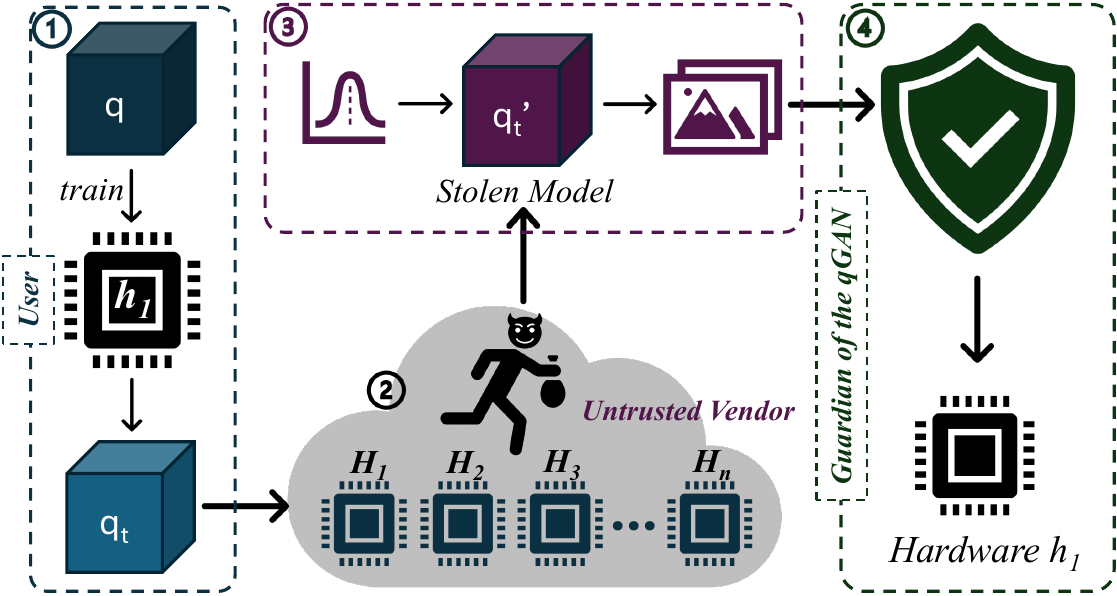}
    \caption{The flow diagram describes our attack model and the proposed security measure. In the figure (1) shows the user training his qGAN, $q$ on hardware $h_1$ to generate a trained qGAN $q_t$; (2), (3) describes the threat model of an untrusted quantum hardware vendor where the user sends $q_t$ for inferencing (note, the hardware used for inferencing, $H_i$, could be different than the hardware used for training $h_1$), from where it gets counterfeited by the untrusted vendor ($q_t^{'}$); (4) is our proposed method of collecting the images generated by $q_t^{'}$ and detecting the hardware where it has been trained using the classifier for proof of ownership.
     }
    \label{fig:flow}
    \vspace{-10pt}
\end{figure}

In the Noisy Intermediate-Scale Quantum (NISQ) era of quantum devices, all quantum computers have noise and the noise inherent to quantum hardware is unique to their configurations in terms of the number of qubits and basis gate sets. Therefore, we propose a watermarking solution to protect the IP of a qGAN based on the following observations of a NISQ \cite{Preskill_2018} device: (i) noise signature from the quantum hardware where the qGAN is trained gets embedded in the images generated. This is evident from Fig. \ref{fig:param} as we observe the influence of noise in the evolution of a parameter in the qGAN under ideal conditions (simulation without noise) and on noisy IBM hardware. Since the noise signature is prominent for different configurations of quantum hardware (number of qubits and noise models), this can be valuable watermark information to detect the hardware where a model has been trained, and (ii) a qGAN when trained on a particular hardware and used to infer images on different hardware, still preserves the original watermark i.e., the hardware on which the model was trained. Let us take an example to explain Fig. \ref{fig:flow} where the user trains the qGAN model ($q$) on a particular hardware ($q_t$) and hosts it on the third-party cloud-based quantum hardware provider for inferencing. However, during the inferencing phase, the trained qGAN model gets stolen ($q_t'$) by the untrusted cloud owner or a rogue adversary sharing the same suite of hardware as the user. The user can claim ownership of the qGAN model by testing a set of images generated by the suspected qGAN model on the classifier to detect the hardware where it was trained. Owing to the robust noise characteristics of different quantum hardware, the classifier detects images generated by qGANs trained on different quantum hardware thus proving the claim of ownership of the original model. \emph{To our knowledge, this is the first attempt to watermark a qGAN circuit}. 
Although we demonstrate the watermarking for qGAN, the idea is equally applicable to other QML models like quantum classifiers. The major contributions are defined as follows:
\begin{enumerate}
    \item Our study utilizes the inherent noise in NISQ-era quantum hardware to watermark a qGAN model.

    \item We perform detailed training of a qGAN on a suite of backends that have noise calibration data from real hardware, varying the noise models and size of the backend. 

    \item We propose to further increase the robustness of the watermark by training the qGAN model on a sequence of multiple hardware to make it difficult to replicate.
    
    \item We train a classical machine learning model to classify the images based on the hardware where the generator model was trained. This would aid in extracting the watermark (i.e., the hardware (or suite of hardware) on which the gGAN was trained) from potentially counterfeit qGAN during the proof of ownership process.

    \item We demonstrate the efficacy of our proposed method by training the qGAN on the digits dataset \cite{misc_optical_recognition_of_handwritten_digits_80} which is widely used for Quantum Machine Learning (QML) evaluation. 

\end{enumerate}

\subsection{Paper Structure}

Section II provides a background for the proposed idea including the model of qGAN that is used for experimentation. Section III goes over the threat model. Section IV presents a detailed description of the proposed idea and includes a comparison of the results. Section V contains an analysis of the efficacy of the idea followed by the security analysis in Section VI and the conclusion in Section VII.
\section{Background}
\subsection{Generative Adversarial Network}
The idea of a generative adversarial network (GAN) is to generate data resembling the original data used in training. This is best understood through the framework of a min-max game, a concept from game theory that involves two players who compete against each other with opposing goals. In the context of GANs, these two players are the generator and the discriminator. The role of the generator is to create synthetic data that mimics the original data while the discriminator tries to classify the data as accurately as a real or a fake. Through iterative training cycles, the generator tries to `fool' the discriminator by improving the quality of the generated images and the discriminator concurrently improves at distinguishing between real and fake data, eventually reaching an equilibrium. The value function for this 2-player min-max game is summarised by, 
\begin{flalign*}
\min_G \max_D V(D, G) = \mathbb{E}_{x \sim p_{\text{data}}(x)}[\log D(x)] \\
+ \mathbb{E}_{z \sim p_z(z)}[\log (1 - D(G(z)))] 
\end{flalign*}
The generator, $G$ starts from some initial latent noise distribution $P_z$ and maps it to $P_g = G(P_z)$. The best solution for this value function would be to achieve $P_g = P_{data}$. 
\subsection{Quantum Generative Adversarial Network}
\begin{figure}
    \centering
    \includegraphics[width=0.7\linewidth]{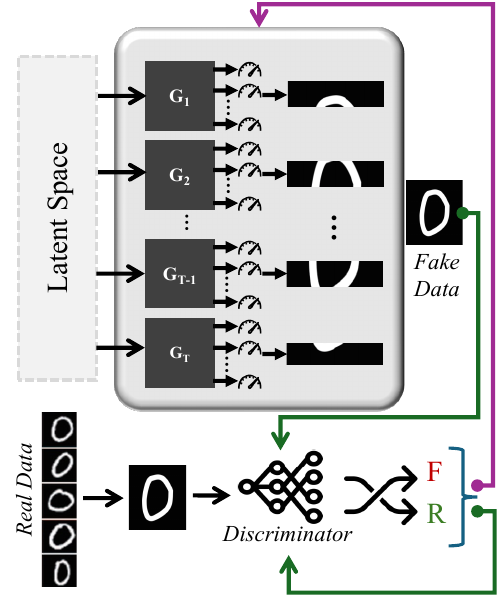}
    \caption{Block diagram of a Patch qGAN model that we implement and train on our suite of IBM backends.
     } 
    \label{fig:patch}
    \vspace{-10pt}
\end{figure}
In this work, we used a Patch QGAN (Fig. \ref{fig:patch}) \cite{Huang_2021} which is designed to operate within the constraints of NISQ machines e.g., a limited number of qubits. The architecture cleverly utilizes a combination of quantum and classical components to generate high-dimensional features using limited quantum resources. This approach allows for the effective use of quantum systems in generating complex data distributions, such as images, by breaking down the task into smaller, manageable segments or ``patches''.
The quantum generator in a Patch QGAN is the core component where quantum computing is employed. It operates based on the following specifics:

\textbf{Sub-Generators Structure:} The generator is composed of multiple sub-generators, each designed as a Parameterized Quantum Circuit (PQC). These sub-generators operate independently to produce segments of the overall data output. This modularity allows the generator to manage high-dimensional data generation tasks by effectively distributing the workload among available quantum resources.

\textbf{Quantum State Preparation:} Each sub-generator receives an encoded quantum state derived from a latent vector sampled from a predefined distribution. This encoding transforms the latent space data into a quantum state that can be manipulated by the quantum circuits.

\textbf{Parameterized Quantum Circuits:} The PQCs employ tunable rotation gates which are adjusted during the training process. These gates manipulate the quantum states to encode the necessary features of the data being generated.

\textbf{Measurement and Assembly:} After processing through the PQCs, the quantum states are measured. The results of these measurements correspond to different features of the output data. These measured outputs from all sub-generators are then assembled to form a complete data instance.

\textbf{Classical Discriminator}
The discriminator in this architecture is entirely classical and functions to evaluate the authenticity of the data generated by the quantum generator. The classical discriminator assesses whether the data instances produced by the generator resemble real data from the training set. It performs binary classification to label data as ``real" or ``fake."

\textbf{Training Process}
The training of Patch QGAN with a classical discriminator follows the typical GAN training protocol but is adapted to accommodate the quantum-classical interaction:

\textit{Forward Pass:} The generator produces data instances, which are forwarded to the discriminator. This step involves transforming quantum measurements into a format compatible with the classical discriminator.

\textit{Loss Evaluation and Parameter Updates:} The discriminator evaluates the generated data against real data and calculates the loss. This loss informs the updates made to the quantum parameters of the generator via quantum-specific optimization techniques, such as quantum gradient estimation methods.

\textit{Iterative Optimization:} The training process iteratively adjusts the parameters of both the generator and the discriminator. For the generator, quantum circuit parameters are tuned to minimize the ability of the discriminator to distinguish generated data from real data. Simultaneously, the discriminator optimizes its parameters to maximize its accuracy in classification.

\begin{table}
\centering
\caption{A comparison of the noise metrics of 127-qubit real and fake quantum hardware }
\begin{tabular}{c||cccc}

\textbf{Hardware} & \textbf{T1 ($\mu s$)} & \textbf{T2 ($\mu s$)} & \textbf{Readout} & \textbf{PauliX} \\ \hline \hline
Brisbane (Fake) & 224.85 & 141.72 & 0.029 & 3.690e-04 \\ 
Kyiv (Fake) & 273.28 & 104.25 & 0.017 & 1.514e-03 \\ 
Osaka (Fake) & 287.31 & 139.96 & 0.042 & 1.357e-03 \\ 
Sherbrook (Fake) & 303.93 & 162.05 & 0.019 & 7.217e-04 \\ \hline \hline
Brisbane (Real) & 222.67 & 151.55 & 0.014 & 2.337e-04 \\ 
Kyiv (Real) & 282.84 & 104.63 & 0.006 & 2.660e-04 \\ 
Osaka (Real) & 259.75 & 121.66 & 0.002 & 3.020e-04 \\  
Sherbrook (Real) & 264.42 & 172.34 & 0.001 & 2.270e-04 \\ 
\hline \hline
\end{tabular}

\label{tab:hardware_metrics}
\end{table}

\subsection{Quantum Noise}
Quantum noise refers to the undesirable disruptions that impact quantum systems, causing errors in quantum computations. These disturbances can originate from multiple sources, such as electromagnetic interference, flaws in quantum gates, thermal variations, and interactions with the environment. \textbf{Gate errors} occur during the implementation of quantum gates. These errors arise from imprecise control or environmental interference affecting the quantum system, leading to incorrect qubit states impacting the accuracy of quantum algorithms. \textbf{Decoherence errors} happen from qubits losing their quantum properties over time due to interactions with their environment which causes the qubits to lose superposition and entanglement leading to the degradation of the quantum state. Decoherence limits the time available to perform quantum operations before the information stored in the qubits becomes unreliable. \textbf{Dephasing errors} occur when the relative phases between the components of a superposition state of a quantum system become imperfections in quantum gate operations. This error type does not affect the probability distribution of the qubit states but disrupts the coherence between them, which can degrade the outcome of quantum algorithms that rely on phase coherence. \textbf{Relaxation error} refers to the process where a qubit in an excited state loses energy and relaxes to a lower energy state more quickly than intended leading to erroneous computational results. Relaxation errors are particularly problematic in maintaining the state of qubits over the duration required for complex quantum calculations. \textbf{Readout error} in quantum computing refers to inaccuracies that occur when measuring the state of qubits at the end of a quantum computation. These errors can arise from technical limitations of the quantum hardware in the measurement process. \textbf{Crosstalk error} in quantum computing occurs when the operation of one qubit inadvertently affects the state of another nearby qubit, due to their physical proximity or electromagnetic interactions. This unwanted interference can alter the intended states of qubits in multi-qubit circuits leading to computational errors.
\subsection{IBM backends}
Qiskit IBM Runtime is a new environment offered by IBM Quantum that streamlines quantum computations and provides optimal implementations of the Qiskit primitives. Several APIs are designed to streamline access to multiple IBM devices. The fake provider module provides a suite of 56 fake backends which are built to mimic the behaviors of the original IBM system snapshots. This has been done to ease the workload on real devices due to the high queue times. To prove the efficacy of our approach we train the qGAN on a suite of different hardware from the IBM fake backends with different coupling maps and noise characteristics. Since the IBM backends are system snapshots, the error values have been calibrated using real hardware as shown in Table \ref{tab:hardware_metrics} (the median values of T1 and T2 times and mean values of readout and PauliX error over all qubits have been shown).

\subsection{Related Work}
Recent research has addressed secure computing on untrustworthy cloud-based quantum hardware providers to some extent. For example, the rise of multi-tenant computing environments in cloud-based quantum hardware and untrusted but efficient quantum compilers has been pointed out as challenging security threats \cite{ghosh2023primer}. Splitting of computation in terms of shots for executing a quantum circuit between trusted and untrusted hardware has been found to be secure, providing $\sim30\%$ times improvement in performance \cite{suryansh_acm}. However, this solution does not work for securing trained qGAN models as splitting the iterations or shots for inferencing the model is not feasible. Another work involves randomly injecting $X$ gates at the last layer of the quantum circuit to obfuscate the quantum circuit and the output from the untrusted cloud. The user can extract the correct output classically by flipping the obfuscated portion \cite{patel2023privacy}. This, again does not work for inferencing qGANs as the adversary could always brute force through the circuit (in reasonable time since the number of qubits and the number of trials to recover the correct qGAN and image will be manageable) to obtain the correct image. There have been attempts to include decoy nodes in the Quantum Approximate Optimization Algorithm (QAOA) circuits to enhance their security and prevent untrusted cloud providers from stealing them \cite{ayanzadeh2023enigma}. An extension of this approach to qGANs by inserting parameterized gates is not useful for securing qGAN models as it only increases the overhead of inferencing without securing the weights of the trained qGAN model. 

There have been approaches to secure the IP of the user by obfuscating them with decoy pulses  \cite{pulse}. The idea is the corrupt the functionality using fake gates/pulses to defeat IP theft. The decoy pulses will be suppressed inside the quantum hardware based on an encrypted classical channel that carries the decoy pulse information. This technique may be applicable to protect the qGAN IP however, it will require the following support from an untrusted cloud provider: (i) a secure classical decrypting hardware inside the quantum software stack to decrypt the pulse map, obtain the decoy gates, and suppress them during execution and, (ii) a pulse suppression feature in the quantum hardware. Both of these features are currently not supported by any quantum cloud provider and will require significant implementation costs. In contrast, the proposed watermarking idea does not require any extra and potentially costly infrastructure from the quantum cloud provider. To summarize, existing work focuses on protecting the IP through obfuscation many of which are not directly applicable to the qGAN domain. Furthermore, watermarking-based solutions, that are covered in this paper, have remained largely unexplored.
\section{Threat Model and analysis}

\subsection{Threat model}
The NISQ-era devices are much in demand due to the increasing complexity of quantum generative models and the increase in their number of parameters. This spikes the demand for quantum hardware to train and infer such models and motivates other third-party vendors to provide execution on quantum hardware as a service. Devices with various qubit technologies, varied numbers of qubits, and varying noise calibrations are typically present in the suite of hardware maintained by a cloud provider. We assume that this quantum hardware is cloud-based and hosted remotely (possibly in untrusted/less trusted countries) which makes it untrustworthy/less trustworthy. 
\begin{table*}
\centering
\caption{Description of the layer sizes and parameters in our CNN used for detecting the training hardware}
\label{tab:cnn_details}
\begin{tabular}{c|c|c|c|c|c|c|c}
\textbf{Layer Type} & \textbf{Output Shape} & \textbf{Kernel Size} & \textbf{Stride} & \textbf{Activation} & \textbf{\# Filters} & \textbf{Pooling Size} & \textbf{Parameters} \\ \hline \hline
Conv2D              & 148x148x32            & 3x3                  & 1               & ReLU                & 32                           & N/A                  & 896                 \\ 
MaxPooling2D        & 74x74x32              & N/A                  & N/A             & N/A                 & N/A                            & 2x2                  & 0                   \\ 
Conv2D              & 72x72x64              & 3x3                  & 1               & ReLU                & 64                           & N/A                  & 18,496              \\ 
MaxPooling2D        & 36x36x64              & N/A                  & N/A             & N/A                 & N/A                           & 2x2                  & 0                   \\ 
Conv2D              & 34x34x128             & 3x3                  & 1               & ReLU                & 128                         & N/A                  & 73,856              \\ 
MaxPooling2D        & 17x17x128             & N/A                  & N/A             & N/A                 & N/A                          & 2x2                  & 0                   \\ 
Flatten             & 36992                 & N/A                  & N/A             & N/A                 & N/A                         & N/A                  & 0                   \\ 
Dense               & 512                   & N/A                  & N/A             & ReLU                & N/A                         & N/A                  & 21,234,176          \\ 
Dense (Output)      & \# hardwares                     & N/A                  & N/A             & Softmax                       & N/A              & N/A                  & 5,130               \\ \hline \hline
\end{tabular}
\end{table*}

Considering that qGAN IP is costly, the untrusted cloud provider or a rogue adversary accessing the same cloud provider as the user may steal the trained qGAN model during inference operation and host the stolen model on some other quantum hardware within their suite or on a different quantum cloud claiming it to be their own. As a result, multiple variants of the same qGAN may be available in the market potentially at a much cheaper price siphoning the profit from the original qGAN provider. 



\subsection{Adversary capability}
We assume that the untrusted cloud provider has: (i) access to the white-box architecture of the qGAN circuit. Therefore, they can strip the noise embedding (i.e., the state preparation circuit) from the qGAN to create a counterfeit copy. Next, they will attach their own noise embedding to this copy which can be offered as their own version of the original qGAN, (ii) the weights of the trained qGAN model. This will require reverse engineering of the transpiled qGAN circuit (where the parametric rotation gates are decomposed into basis gates) to recover the pre-transpiled version of the trained qGAN circuit. This ability will allow the adversary to perform further training to potentially tamper with the watermark.

\subsection{Assumptions}
We assume as a part of the threat model that the hardware suite used by the user for training the qGAN is different from the one used to inference the qGAN models on the untrusted cloud.
\section{Proposed Idea}


\subsection{Architecture of quantum GAN}
\begin{figure}
    \centering
    \includegraphics[width=0.8\linewidth]{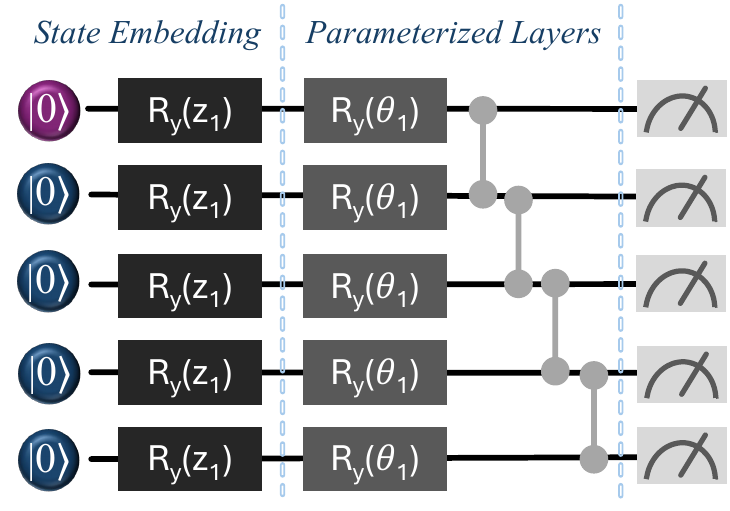}
    \caption{A Parameterized quantum circuit design for the sub-generator to generate a patch for the fake image. The purple qubit represents the ancilla qubit and the blue qubits represent the data qubits.
     }
    \label{fig:qgan_ckt}
    \vspace{-10pt}
\end{figure}

Our approach is recreated from \cite{Huang_2021} and utilizes a series of quantum generators, where each sub-generator, denoted \( G(i) \), (Fig. \ref{fig:qgan_ckt}) is assigned the task of forming a distinct segment of the final image. These segments are subsequently concatenated to assemble the complete image, as illustrated below. One of the primary advantages of our methodology is its adaptability to scenarios with limited qubit availability. The same quantum device can be employed iteratively for each sub-generator, or the sub-generators can be executed in parallel across multiple devices. \\
\subsubsection{Implementing the Generator}

Our approach involves a network of sub-generators, each denoted as \(G(i)\), which share a consistent circuit architecture as depicted below. The complete quantum generator comprises \(N_G\) sub-generators, each equipped with \(N\) qubits. The transformation process from the input of a latent vector to the final image output is divided into four key stages: state embedding, parameterization, non-linear transformation, and post-processing. To clarify, each of the sections described below corresponds to a single cycle of the training process.

\begin{enumerate}
    \item \textbf{State Embedding:}
    A latent vector \(z \in \mathbb{R}^N\) is randomly selected from a uniform distribution within the range \([0, \pi/2)\). This vector is simultaneously fed into all sub-generators, where it is embedded using RY gates.

    \item \textbf{Parameterized Layers:}
    This stage involves parameterized RY gates followed by controlled Z gates, repeated \(D\) times in sequence.

    \item \textbf{Non-Linear Transformation:}
    Given that quantum gates are inherently unitary and thus linear, our challenge is to induce non-linear transformations for complex generative tasks. We employ ancillary qubits for this purpose. For any sub-generator, the quantum state prior to measurement is described by:
    \[
    |\Psi(z)\rangle = U_G(\theta)|z\rangle
    \]
    Post-measurement, when tracing out the ancillary subsystem \(A\), the state \(\rho(z)\) becomes:
    \[
    \begin{split}
    \rho(z) & = \frac{\text{Tr}_A(\Pi \otimes I|\Psi(z)\rangle\langle\Psi(z)|)}{\text{Tr}(\Pi \otimes I|\Psi(z)\rangle\langle\Psi(z)|)} \\ & = \frac{\text{Tr}_A(\Pi \otimes I|\Psi(z)\rangle\langle\Psi(z)|)}{\langle\Psi(z)|\Pi \otimes I |\Psi(z)\rangle}        
    \end{split}
    \]
    The non-linear transformation is evidenced by the dependency of \(\rho(z)\) on \(z\) in both the numerator and denominator.

    \item \textbf{Post Processing:}
    The measured probability \(\rho(z)\) of each computational basis state \(P(j)\) is used to derive the output \(g(i)\) from the sub-generator:
    \[
    g(i) = [P(0), P(1), \ldots, P(2^{N - N_A} - 1)]
    \]
    Due to measurement normalization constraints, where the sum of all elements in \(g(i)\) must equal one, we apply a post-processing normalization to convert \(g(i)\) into usable pixel intensity values for image patches:
    \[
    \tilde{x}(i) = \frac{g(i)}{\max_k g(i)_k}
    \]
    Thus, the final constructed image \(\tilde{x}\) is formed by assembling these normalized patches:
    \[
    \tilde{x} = [\tilde{x}(1), \ldots, \tilde{x}(N_G)]
    \]
\end{enumerate}
\subsubsection{Implementing the Discriminator}
The Discriminator is structured as a fully connected classical neural network within the PyTorch framework, designed with two hidden layers to classify images as real or generated. It takes a flattened image as input, processed through layers of 64 and 16 neurons respectively, each followed by ReLU activations. The output is a single neuron with a sigmoid activation function, yielding a probability score. The network architecture enables it to learn and make binary classifications during GAN training. 



\subsection{Watermark generation and extraction}
The core property of NISQ-era quantum hardware is its noisy behavior. We note that the model parameters evolve differently under quantum noise during training. As such, training of the qGAN on a particular quantum hardware with specific noise embeds the watermark naturally within the model parameters. Therefore, the generated images are different than the images generated by a qGAN model which is trained under noiseless conditions or on a different hardware. The watermark will persist even after the adversary steals and reuses the model. During the proof of ownership, the images obtained from the stolen model will be analyzed to extract the watermark i.e., the hardware that was used for training. If the watermark matches with the claim, the proof of ownership will be established.  
For the watermark extraction, we designed a classifier that can classify the images generated by the suspected qGAN to detect the quantum hardware where the qGAN had been trained. The proof of ownership relies on robust classification of the generated images. Therefore, we create a dataset by training a qGAN on various hardware and inferring on a suite of hardware.  


\subsection{Classifier for watermark extraction}
To amplify the features of the images used for training the classifier, we increase the resolution from 8x8 to 150x150 to help the classifier learn better. The architecture comprises an initial input layer that accepts images of dimensions 150x150 pixels, followed by a series of three convolutional layers with 32, 64, and 128 filters of stride size 3x3 respectively. Each convolutional layer utilizes the ReLU activation function to introduce non-linearity. These layers are interspersed with MaxPooling layers of size 2x2 with a stride of 2, aimed at reducing the dimensionality of the feature maps. This mitigates the risk of overfitting the model.

Subsequent to the Convolutional and MaxPooling layers, the network transitions to a fully connected segment through a flattening operation that transforms the spatial feature maps into a dense vector. The vector feeds into a densely connected layer of 512 neurons, also activated by ReLU, which serves to integrate learned features across the image. The final layer of the network is a softmax layer that maps the output of the previous dense layer to a probability distribution over the class labels. This layer has dimensions based on the number of hardware in the suite and the number of labels representing images generated from other hardware. More details about the classifier architecture are present in Table \ref{tab:cnn_details}.

\begin{table}
\caption{Noise characteristics for our suite of IBM backends}
\centering
\begin{tabular}{cc||ccc}
\textbf{Hardware} & \textbf{\# Qubits} & \textbf{Readout} & \textbf{Pauli-X} & \textbf{TR} \\ \hline \hline
ibm\_athens & 5 & 0.017 & 4.82e-04 & 0.035 \\ 
ibm\_bogota & 5 & 0.038 & 4.00e-04 & 0.019 \\ 
ibm\_burlington & 5 & 0.035 & 7.02e-04 & 0.027 \\ 
ibm\_jakarta & 7 & 0.025 & 3.49e-04 & 0.041 \\ 
ibm\_nairobi & 7 & 0.027 & 3.06e-04 & 0.021 \\ 
ibm\_lagos & 7 & 0.009 & 2.58e-04 & 0.023 \\ 
ibm\_cairo & 27 & 0.016 & 3.07e-04 & 0.024 \\ 
ibm\_cambridge & 27 & 0.107 & 9.59e-04 & 0.039 \\ 
ibm\_kolkata & 27 & 0.012 & 3.20e-04 & 0.022 \\ 
ibm\_washington & 127 & 0.049 & 2.00e-04 & 0.030 \\ \hline \hline
\end{tabular}
\label{tab:quantum_devices}
\end{table}

\subsection{Training}
To provide a robust comparison for our watermark extraction method, we consider two use cases for training the qGAN model. First, we consider that the user trains the qGAN on a single quantum hardware till near-final stability and hosts the trained qGAN model on an untrusted quantum hardware provider. In this case, the weights of the trained qGAN model are tuned to the noise of the single quantum hardware where the qGAN was trained. Second, we consider that the user trains the qGAN on a particular hardware for a few epochs and then migrates the partially trained model to different hardware. This route of training quantum machine learning models is observed to escape longer queue times of third-party quantum hardware. In such a situation, the trained qGAN model will have weights that possess the noise characteristics of multiple quantum hardware.

\subsection{Validating ownership}
Since the watermark extraction classifier trains on the data of a known suite of quantum hardware, it will be incapable of detecting images from a qGAN trained on unknown quantum hardware. This raises the concern of misclassifying an original qGAN model as stolen. To encounter this situation, we define a threshold value $\mathcal{M}$ which determines the probability with which the classifier detects the quantum hardware. Owing to the training and validation set of images and the number of parameters and complexity of the qGAN being trained, we consider $\mathcal{M}$ to be a hyperparameter decided by the user to claim his ownership of the model. More details about $\mathcal{M}$ have been discussed with results in the next section.


\section{Results}

This section describes the experimental results and discusses the efficacy of the proposed idea.
\subsection{Simulation setup}
\textbf{Training: }The qGAN model has been implemented in Pennylane \cite{bergholm2018pennylane} using PyTorch \cite{paszke2019pytorch} as a wrapper for ease of pipelining the flow. The sub-generators of the qGAN use 5 qubit parameterized quantum circuits (PQCs) for generating the fake image. We implement 4 sub-generators to generate the patches which totals 100 tunable parameters for the qGAN, and train the qGAN for 500 epochs before inferencing the trained model on multiple quantum hardware. The SGD (Standard Gradient Descent) optimizer is used for both the quantum generator and the discriminator and a BCE (Binary Cross Entropy) loss is the shared loss of both models. The learning rate for the discriminator is set to 0.01 which is quite less than that of the quantum generator (=0.2). The classifier for detecting the quantum hardware is implemented using TensorFlow \cite{tensorflow2015-whitepaper} with a batch size of 64. We use the Adam optimizer and Categorical Crossentropy to evaluate the accuracy of our proposed model within 15 epochs. \textbf{Dataset: } Since we are training our qGAN on IBM backends with limited qubits, high error rates, and complex data encoding, we conduct all experiments using the handwritten digit dataset \cite{misc_optical_recognition_of_handwritten_digits_80} as a proof of concept for our proposed model. Each training sample is an 8x8-pixel image. \textbf{IBM backends: }We run our simulations on the open-source quantum software development kit from IBM (Qiskit). To validate the proposed algorithm we pick a suite of 10 IBM backends shown in Table \ref{tab:quantum_devices}. Here we consider the mean values of readout, PauliX, and thermal relaxation (TR) error over all qubits to train the qGAN model of the user and infer it to produce a rich dataset for training the classifier.

\subsection{Image quality}

\begin{figure}
    \centering
    \includegraphics[width=\linewidth]{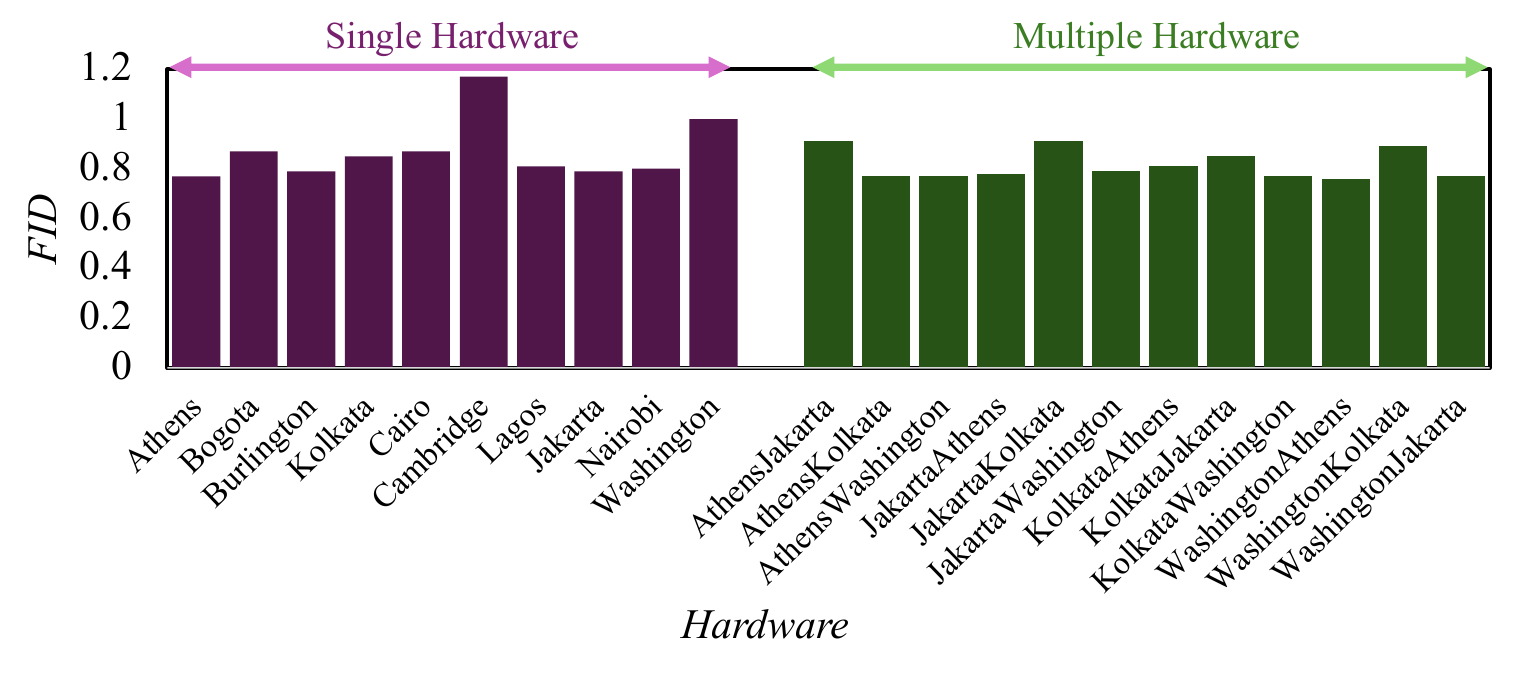}
    \caption{Plot demonstrating the image quality using FID score for images generated by the qGAN when trained on a single hardware and on multiple hardware.}
     
    \label{fig:fid}
    \vspace{-10pt}
\end{figure}

\begin{figure}
    \centering
    \includegraphics[width=\linewidth]{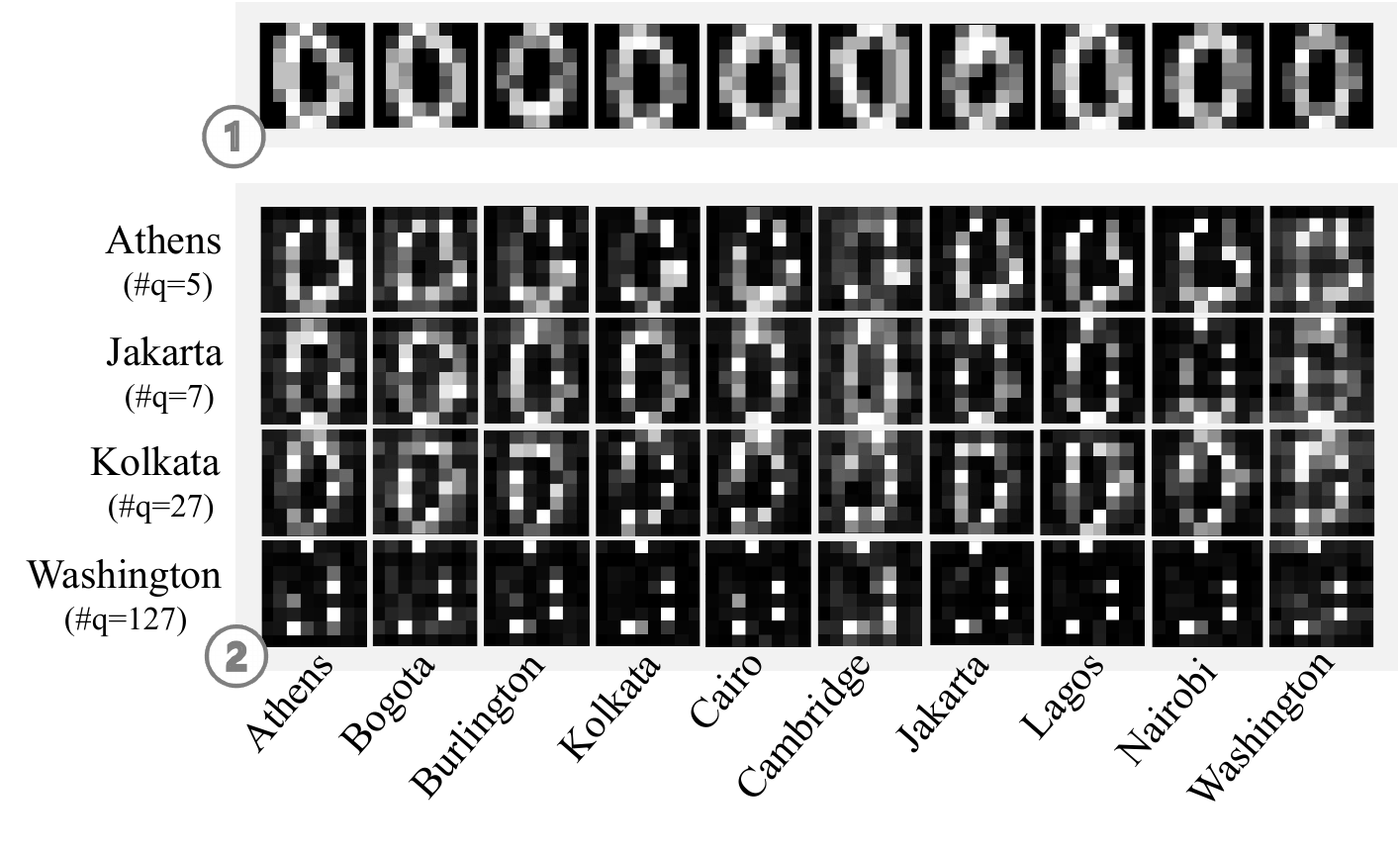}
    \caption{In this figure (1) shows the original digits dataset with label 0 and (2) shows the images generated qGANs trained on different hardware (along the rows) and inference on 10 IBM backends (along the columns). It can be observed that the noise watermark of the training hardware gets embedded in the images making them visually distinct.
     }
    \label{fig:box}
    \vspace{-10pt}
\end{figure}

We use the Fr\'echet Inception Distance (FID) score \cite{heusel2018gans} to compare the quality of generated images with the images of the original dataset. FID score is used as an evaluation metric based on the distance between two Gaussian distributions. For two distributions with mean $(m_1, C_1)$ and mean $(m_2, C_2)$, the $FID = ||m_1-m_2||_2^2+Tr(C_1+C_2-2(C_1C_2)^{1/2})$. A lower FID score between two distributions implies higher similarity. Fig. \ref{fig:fid} compares the FID values of the images from the original dataset and the images generated by the qGAN trained by the user on a single hardware and on multiple hardware. It can be observed that, in spite of the inherent noise in the IBM backend, the generated images have a relatively lower FID score determining their similarity with the original images. For multiple hardware training, the user might be concerned about degraded quality due to two different noise models. However, that is not the case as seen in Fig. \ref{fig:fid}. We describe a subset of images generated in Fig. \ref{fig:box}. We can observe that the images when generated by qGANs trained on different IBM backends, show distinct pixel characteristics which machine learning models can learn to help detect the hardware the qGAN has been trained on.

\begin{figure}
    \centering
    \includegraphics[width=\linewidth]{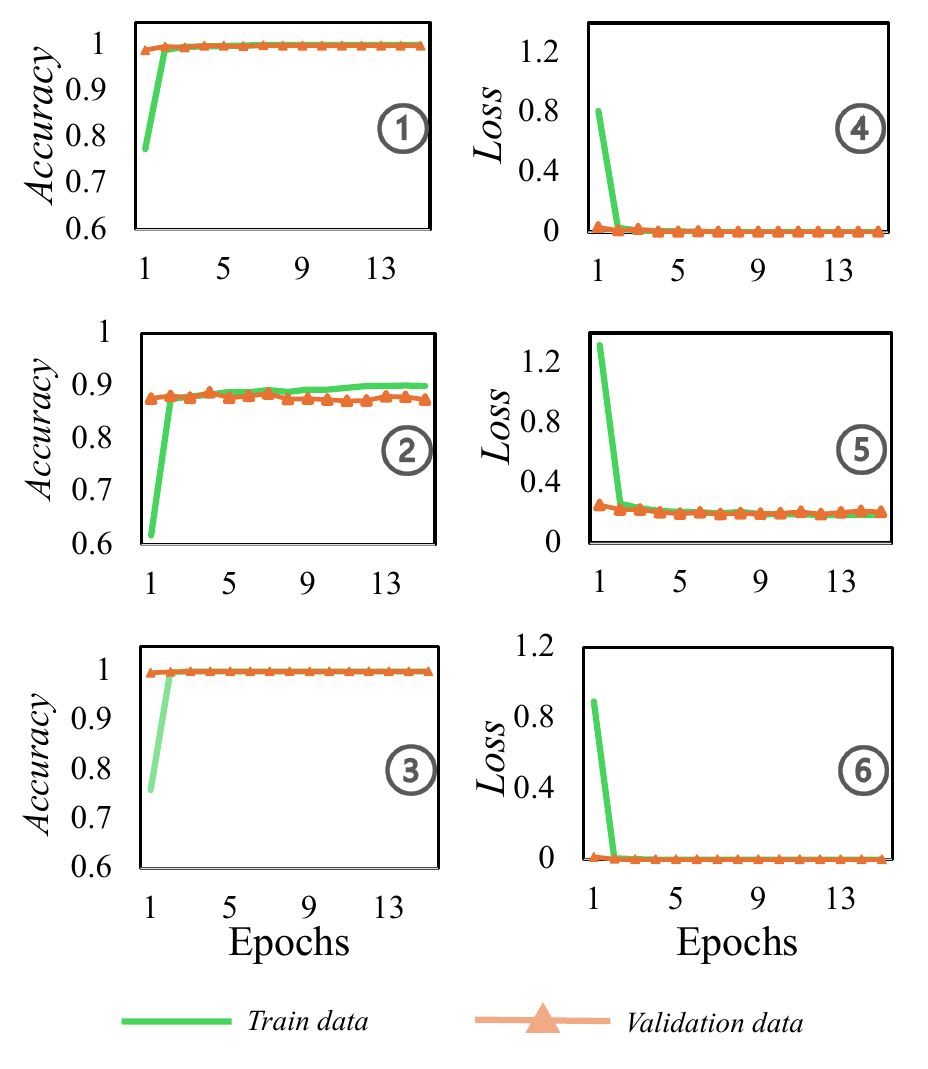}
    \caption{Plot describing the high accuracy of the classifier when trained on images generated by qGANs trained on different hardware. 1, 2 shows the accuracy and loss for single hardware training. 3, 4 shows the accuracy and loss for multiple hardware training. 5, 6 shows the accuracy and loss for a case where we test the impact of fixing the order of hardware during the training of the qGAN.
     }
    \label{fig:classifier}
    \vspace{-10pt}
\end{figure}

\subsection{Watermark extraction}

\begin{figure}
    \centering
    \includegraphics[width=\linewidth]{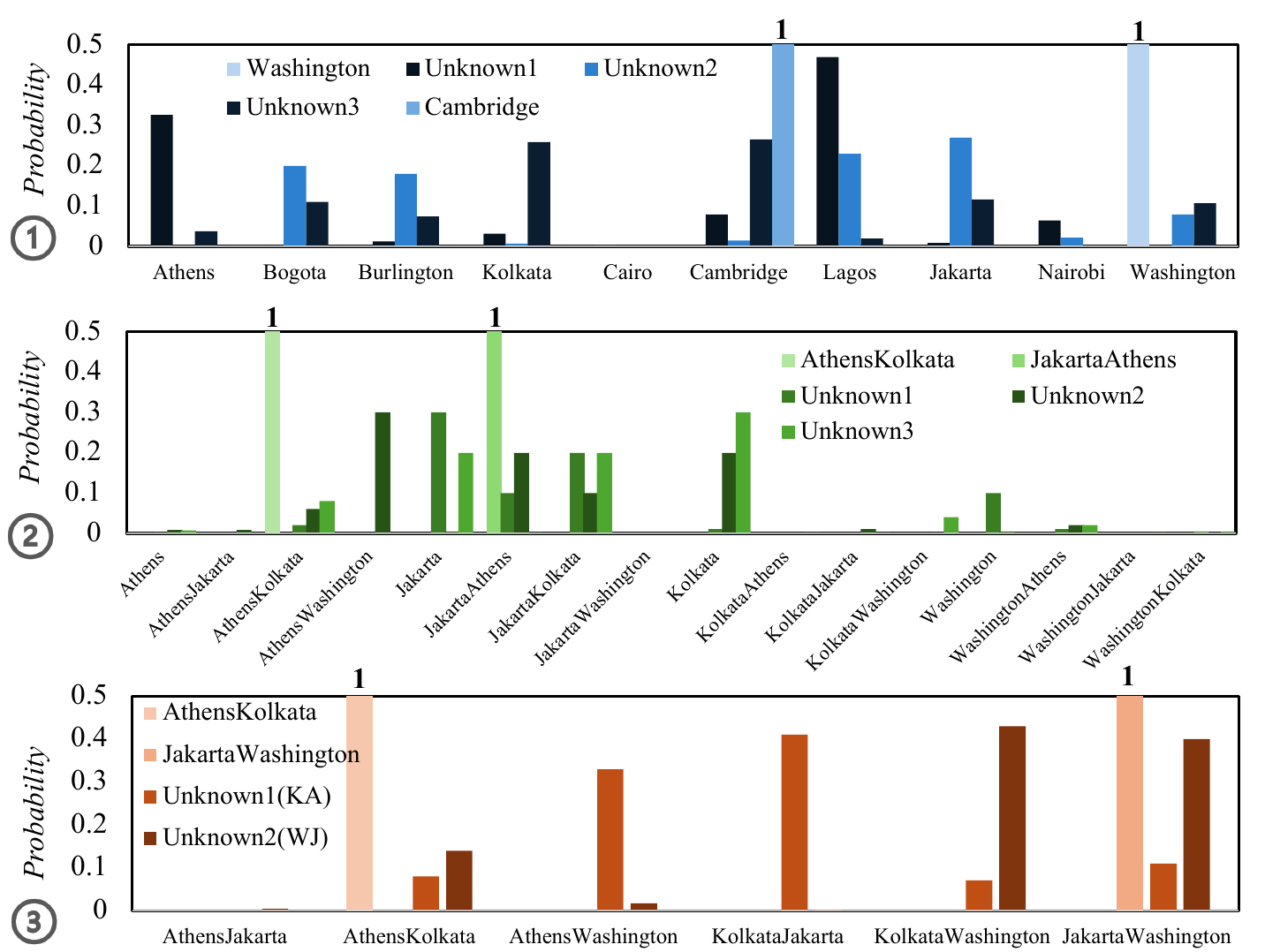}
    \caption{In these plots we describe the high predicted probability for known hardware while validating the proof of ownership using the classifier. The classifier is trained to classify images generated by the qGAN trained on the hardware marked on the x-axis. Figure (1) demonstrates the validation during the training of the qGAN on a single quantum hardware, (2) demonstrates the validation during the training of the qGAN on a sequence of quantum hardware; (3) demonstrates the prominence of the sequence of hardware on which the qGAN is trained.
     }
    \label{fig:singletest}
    \vspace{-10pt}
\end{figure}

\textbf{Extraction for Single Hardware Training: }We consider that the user has trained the qGAN on a single hardware. To train the classifier, we generate a dataset from different IBM backends. We train the qGAN model on all 10 IBM backends and generate the images while inferencing on the backends. We generate 100 images every time the trained model is inference on a different IBM backend. This way we obtain a rich dataset of images having 10 labels representing the hardware where the qGAN was trained. From Fig \ref{fig:classifier} (1, 3) we can observe that the classifier trains with near-perfect accuracy. Owing to the richness of the dataset produced due to the influence of noise of the IBM backends, the task of earning patterns in the data is not a very complex task. To make the extraction task very efficient, we set a threshold value, $\mathcal{M}$. To define $\mathcal{M}$, we test the classifier with a test dataset comprising images generated by the trained qGAN on different IBM backends and set $\mathcal{M}$ as the mean of the probability of the detected training hardware by the classifier for each image. To ensure that the test and training datasets differ, we re-train the qGAN and re-generate the images using random seed values. For the situation where the user trains on a single hardware, we obtain $\mathcal{M}=0.9999$ for our qGAN model and dataset. We demonstrate a subset of the results in Fig \ref{fig:singletest}.1, where we test the trained classifier using a test dataset consisting of 5 labels, each having 1000 images. We construct the test dataset using images generated from qGANs trained on two quantum hardware in our suite (IBM Washington and IBM Cambridge), and three not included in our suite (Unknown1-IBM Casablanca, Unknown2-IBM Almaden, Unknown3-IBM Essex). We can observe that the images generated from known backends are classified correctly with a probability of $\sim1$. Although the images generated from unknown sources are classified as known hardware, the probability of predicting the correct label of hardware is less than the value of $\mathcal{M}(=0.9999)$, which proves our case of handling cases where the suspected qGAN models belong to the user (known hardware) or not (unknown hardware).

\textbf{Extraction for Multiple Hardware Training: }
To demonstrate this procedure we pick 4 IBM backends from our suite having 5 (IBM Athens), 7 (IBM Jakarta), 27 (IBM Kolkata), and 127 (IBM Washington) qubits and train the qGAN on all possible combinations taking two at a time providing a set of 12 labels of known hardware. We also include the dataset of images generated from the single hardware whose combinations we provide, making a final dataset of 16,000 images generated from the qGAN model trained on 12 combinations of 4 IBM backends and the 4 single backends, to train the classifier. We observe from Fig. \ref{fig:classifier}(2,4) that the classifier trains with a reduced accuracy of around 90\%. Owing to the multiple noise models influencing the generated images, the quality of the dataset used to train the classifier was reduced slightly, therefore decreasing the accuracy of the prediction. We find that while testing the trained classifier with a dataset of images generated from known combinations of hardware, the mean probability of detecting the correct hardware combination is almost equal to 1($\sim0.9$) which becomes our threshold value, $\mathcal{M}$. We describe a subset of these results in Fig. \ref{fig:singletest}.2 where we test the classifier on a dataset having 5 labels, each with 1000 images generated similarly as the first use case to ensure that the classifier has never seen the test data. The test dataset has images generated by qGANs trained on known quantum hardware (a sequence of IBM Athens and IBM Kolkata, and a sequence of IBM Jakarta and IBM Athens), and three unknown quantum hardware (Unknown1-IBM Casablanca, Unknown2-IBM Almaden, Unknown3-IBM Essex). We observe that the images generated from the qGAN trained on known hardware combinations get predicted with a probability of $\sim1$. Although the images generated from unknown sources are classified as known hardware combinations, the probability of predicting the correct label of hardware is less than the value of $\mathcal{M}(=0.9)$, which proves our case of handling cases where the suspected qGAN models belong to the user (known hardware) or not (unknown hardware).
We also counter a corner case where the user trains the qGAN model on a set of quantum hardware in a particular sequence, and the suspected qGAN model has been trained on the same set of quantum hardware but in a different sequence. We validate our results in this case by training the classifier on a reduced dataset from the sub-section of model migration. We train the qGAN model on unique combinations from the 4 chosen IBM backends, taking 2 at a time. We can observe from Fig. \ref{fig:classifier} that the classifier in this situation trains with almost 100\% accuracy. We proceed to calculate the threshold value ($\mathcal{M}$) to be $0.999$. To test the case, we prepare a dataset comprising 4 labels, each with 1000 images generated by the qGAN trained on two known combinations (a sequence of IBM Athens and IBM Kolkata, and a sequence of IBM Jakarta and IBM Washington) in two unknown combinations (a sequence of IBM Kolkata and IBM Athens, and a sequence of IBM Washington and IBM Jakarta). Fig. \ref{fig:singletest}.3 shows that the known hardware combinations are predicted with a near-perfect probability and the unknown sequence of hardware gets classified with a probability much less than the threshold value, $\mathcal{M}(=0.999)$. This proves that the noise watermark embedded in the images generated by the qGAN depends on the exact sequence of the hardware on which the qGAN is trained.

\subsection{Overhead Analysis}
\begin{figure}
    \centering
    \includegraphics[width=\linewidth]{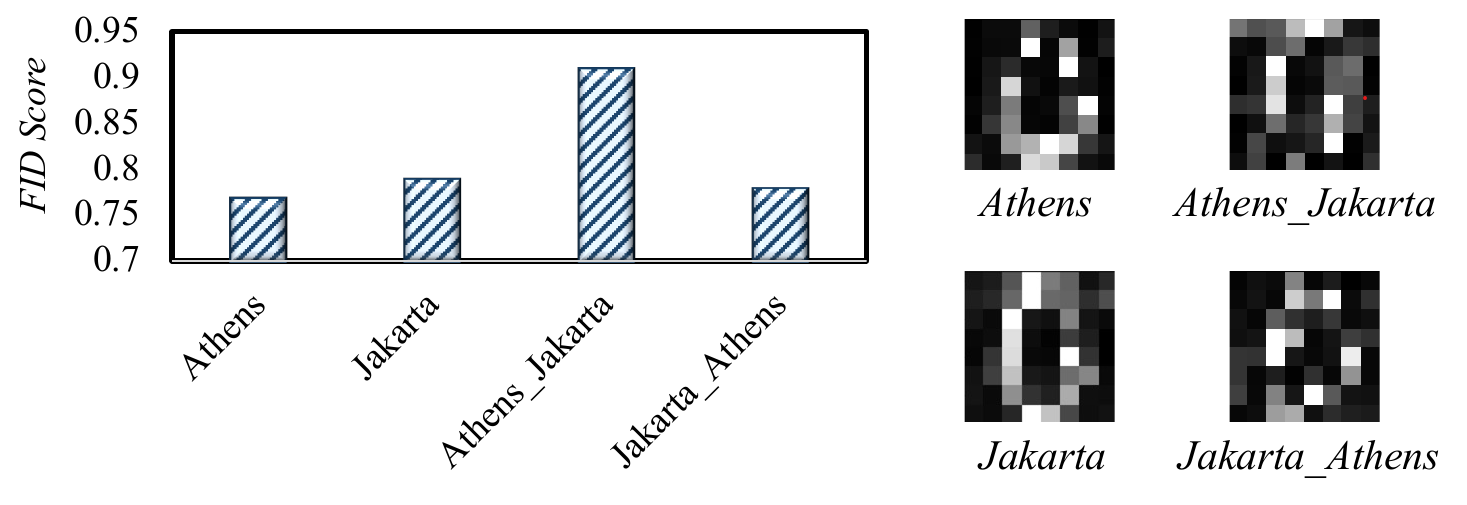}
    \caption{ In this figure, (1) represents the difference of FID scores between images generated by the qGAN trained on IBM Athens, IBM Jakarta, and a combination of both. The FID scores are taken on a mean of 1000 images. (2) A visual representation of the images generated on the same.} 

    \label{fig:fid_anal}
    \vspace{-10pt}
\end{figure}
We present an analysis of the overhead incurred to embed the watermark during the training of the qGAN. Since the user trains the qGAN on noisy hardware, the influence of noise on the weights of the trained model will occur naturally without any overhead. However, when the user shifts the training to a sequence of models, there will be an influence of multiple noise signatures on the images generated by the qGAN which may degrade the image quality. From Fig. \ref{fig:fid_anal} we observe that the FID score for images generated on a sequence of IBM Athens and IBM Jakarta is $\sim 18\%$ higher than the images generated after individual training on each of the devices. However, images generated by a qGAN trained on a sequence of IBM Jakarta followed by IBM Athens are almost identical in terms of FID score. On the entire dataset, the FID scores have $\sim 7.4\%$ increase for the images generated by the qGAN trained on a sequence of hardware than on single hardware. Visually, the images are distinct (Fig. \ref{fig:fid_anal}) and have sufficient features for the classifier to learn and extract the noise watermark in terms of the hardware (or sequence of hardware) where the qGAN was trained. 

\section{Security Analysis and Discussions}
\subsection{Security Analysis}

\subsubsection{Uniqueness of the watermark}
Consider a suite of $n$ quantum hardware on which the user can train the qGAN model. If the user decides to train the qGAN model on a single hardware the probability of collision with another user training a similar qGAN model will be $1/n$. To increase the robustness of the watermark in the trained qGAN model of the user, if the training is done on a sequence of $k$ quantum hardware out of $n$, the probability of collision reduces to $\prod_{i=1}^{k} 1/(n-i)$. In our experimental setup, the probability of collision for a single hardware training is $0.1$ which gets lowered by $\sim13$ times when the user shifts to training on a sequence of two quantum hardware. The chances of collision and proof of ownership could be further boosted by increasing the number of hardware used for training. For instance, IBM has a suite of 15 cloud-based quantum hardware. If the user chooses a sequence of 5 hardware for training, the probability of collision of two users with the same set of hardware in the same sequence will be $\sim10^{-5}$, proving the uniqueness of the watermark further. 


\subsubsection{Removal of watermark}
The watermark in the images generated by the qGAN is directly dependent on the noise model of the quantum hardware where it was trained. Even if we assume that the rogue adversary has a reverse engineering toolkit to obtain the original quantum circuit with the same rotation values as set by the user from the transpiled intermediate, it is computationally costly to un-learn the weights of the trained qGAN \cite{DBLP:journals/corr/abs-1912-03817} owing to the complex stochasticity in the training process and limited knowledge about the initial data points to remove the watermark and ultimately reduces the incentive of the adversary to steal the qGAN model.

\subsubsection{Tampering the watermark}
\begin{figure}
    \centering
    \includegraphics[width=0.8\linewidth]{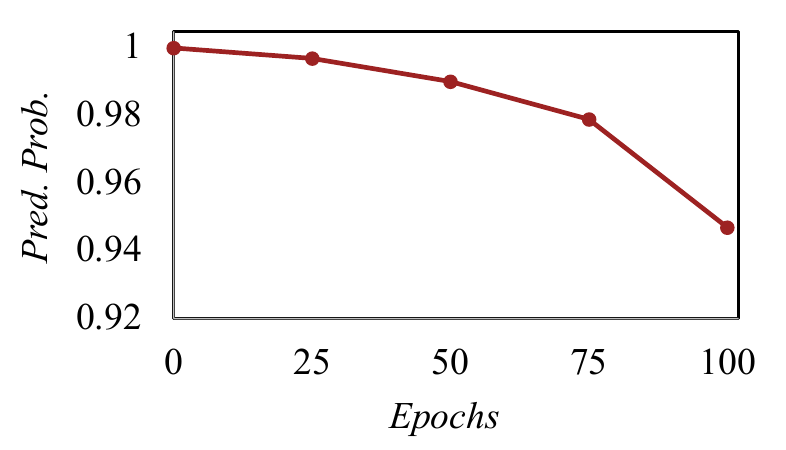}
    \caption{ Plot describing the reduction in prediction probability after tampering with the watermark by training on unknown hardware. The original qGAN model was trained on a sequence of IBM Athens and IBM Jakarta. 
     }
    \label{fig:tampering}
    \vspace{-10pt}
\end{figure}
There might be attempts by an untrusted vendor or rogue adversary to alter the watermark in a stolen qGAN model to derail the proof of ownership. Suppose the adversary has access to the trained qGAN model and the tools to retrieve the original quantum circuit used by the owner. They could try to overwrite the original watermark by training the stolen model on different, unknown quantum hardware for a short period. This approach will minimize the cost of extra training. To evaluate our security measures against such tampering, we trained a qGAN model on a known sequence of quantum hardware (IBM Athens and IBM Jakarta), then briefly trained it on different hardware, and conducted inference tests on various hardware. The results, shown in Fig. \ref{fig:tampering}, indicate that our classifier can still identify the original hardware sequence correctly even after the model was trained on different hardware for a few epochs, with a high level of prediction probability ($>\mathcal{M}$). However, completely altering the watermark would require much more extensive retraining, which is unlikely to be cost-effective, reducing the motivation to steal and tamper with the model.

\subsubsection{Ghost watermark}
When an adversary falsely claims their watermark exists in a stolen qGAN model based on some pattern in the images generated by the stolen qGAN, in the absence of a real watermark, it will be referred to as a ghost watermark \cite{ghost}. The possibility of claiming a ghost watermark depends on how cleverly the adversary can mimic a feature hidden in the trained qGAN model. One can defeat ghost watermarks by including multiple watermarks in the trained qGAN which will offer stronger proof of ownership. This could be a topic for further research.

\subsection{Discussion}
\subsubsection{Watermark of other QML models}
The watermark extraction procedure and accuracy on multiple hardware are high enough to conclude the influence of quantum hardware noise in the training of generic QML models and hence the proposed method could be extended to other QML circuits like Quantum classifiers by observing the pattern of the probability vectors generated during the classification.

\subsubsection{Hardware validation}
Due to large queue times for training and inferencing large QML models like qGAN on cloud-based quantum hardware provided by companies like IBM, we decided to provide extensive results by simulating our proposed setup on IBM backends that are snapshots of original IBM quantum hardware calibrated with the original noise values (Table \ref{tab:hardware_metrics}) to provide a proof-of-concept for our proposed idea. Recent research in the development of quantum machine learning applications like Variational Quantum Circuits \cite{robbiati2023realtime}, and quantum approximate estimation using Variational Quantum Eigensolvers \cite{Sack_2024} show that the parameters and loss function evolve differently on real quantum hardware compared to noiseless simulations. This confirms our idea that quantum noise embeds hardware noise-specific watermarks (i.e., different evolution of parameters) during training on real hardware. Therefore, the ideas and conclusions drawn in this paper will remain the same even on real hardware. 





\section{Conclusion}
This paper leverages the unique noise characteristics of the quantum hardware used for training to watermark quantum generative adversarial networks (qGANs) for securing intellectual property (IP) against unauthorized use and theft by third-party quantum computing resources. We note that noise in quantum hardware gets embedded in the model parameters as a footprint which in turn is reflected in the generated images by the qGANs. Tested across ten IBM quantum backends, our method demonstrated a 100\% accuracy rate in extracting the watermark from the images generated by qGANs trained on a single hardware and $\sim90\%$ accuracy rate when the qGAN was trained on a sequence of quantum hardware to incorporate a more complex watermark. The security analysis confirms the resilience of the proposed watermark against removal and tampering. 

\section*{Acknowledgment}

The work is supported in parts by the National Science Foundation (NSF) (CNS-1722557, CCF-1718474, OIA-2040667, DGE-1723687, and DGE-1821766) and gifts from Intel.

\bibliographystyle{unsrt}
\bibliography{refs}

\begin{thebibliography}{10}

\bibitem{gan1}
Ian~J. Goodfellow, Jean Pouget-Abadie, Mehdi Mirza, Bing Xu, David Warde-Farley, Sherjil Ozair, Aaron Courville, and Yoshua Bengio.
\newblock Generative adversarial networks, 2014.

\bibitem{hu2022text}
Kai Hu, Wentong Liao, Michael~Ying Yang, and Bodo Rosenhahn.
\newblock Text to image generation with semantic-spatial aware gan, 2022.

\bibitem{ledig2017photorealistic}
Christian Ledig, Lucas Theis, Ferenc Huszar, Jose Caballero, Andrew Cunningham, Alejandro Acosta, Andrew Aitken, Alykhan Tejani, Johannes Totz, Zehan Wang, and Wenzhe Shi.
\newblock Photo-realistic single image super-resolution using a generative adversarial network, 2017.

\bibitem{wang2022highfidelity}
Tengfei Wang, Yong Zhang, Yanbo Fan, Jue Wang, and Qifeng Chen.
\newblock High-fidelity gan inversion for image attribute editing, 2022.

\bibitem{Dallaire_Demers_2018}
Pierre-Luc Dallaire-Demers and Nathan Killoran.
\newblock Quantum generative adversarial networks.
\newblock {\em Physical Review A}, 98(1), July 2018.

\bibitem{sharma2022analysis}
Aakash Sharma, Vivek~M. Bhasi, Sonali Singh, Rishabh Jain, Jashwant~Raj Gunasekaran, Subrata Mitra, Mahmut~Taylan Kandemir, George Kesidis, and Chita~R. Das.
\newblock Analysis of distributed deep learning in the cloud, 2022.

\bibitem{upadhyay2023trustworthy}
Suryansh Upadhyay, Rasit~Onur Topaloglu, and Swaroop Ghosh.
\newblock Trustworthy computing using untrusted cloud-based quantum hardware, 2023.

\bibitem{szyller2023good}
Sebastian Szyller, Vasisht Duddu, Tommi Gröndahl, and N.~Asokan.
\newblock Good artists copy, great artists steal: Model extraction attacks against image translation models, 2023.

\bibitem{Yuan2022}
C.~Yuan, H.~Wang, P.~He, et~al.
\newblock Gan-based image steganography for enhancing security via adversarial attack and pixel-wise deep fusion.
\newblock {\em Multimedia Tools and Applications}, 81:6681--6701, 2022.

\bibitem{Preskill_2018}
John Preskill.
\newblock Quantum computing in the nisq era and beyond.
\newblock {\em Quantum}, 2:79, August 2018.

\bibitem{misc_optical_recognition_of_handwritten_digits_80}
E.~Alpaydin and C.~Kaynak.
\newblock {Optical Recognition of Handwritten Digits}.
\newblock UCI Machine Learning Repository, 1998.
\newblock {DOI}: https://doi.org/10.24432/C50P49.

\bibitem{Huang_2021}
He-Liang Huang, Yuxuan Du, Ming Gong, Youwei Zhao, Yulin Wu, Chaoyue Wang, Shaowei Li, Futian Liang, Jin Lin, Yu~Xu, Rui Yang, Tongliang Liu, Min-Hsiu Hsieh, Hui Deng, Hao Rong, Cheng-Zhi Peng, Chao-Yang Lu, Yu-Ao Chen, Dacheng Tao, Xiaobo Zhu, and Jian-Wei Pan.
\newblock Experimental quantum generative adversarial networks for image generation.
\newblock {\em Physical Review Applied}, 16(2), August 2021.

\bibitem{ghosh2023primer}
Swaroop Ghosh, Suryansh Upadhyay, and Abdullah~Ash Saki.
\newblock A primer on security of quantum computing, 2023.

\bibitem{suryansh_acm}
Suryansh Upadhyay and Swaroop Ghosh.
\newblock Robust and secure hybrid quantum-classical computation on untrusted cloud-based quantum hardware.
\newblock In {\em Proceedings of the 11th International Workshop on Hardware and Architectural Support for Security and Privacy}, HASP '22, page 45–52, New York, NY, USA, 2023. Association for Computing Machinery.

\bibitem{patel2023privacy}
Tirthak Patel, Daniel Silver, Aditya Ranjan, Harshitta Gandhi, William Cutler, and Devesh Tiwari.
\newblock Toward privacy in quantum program execution on untrusted quantum cloud computing machines for business-sensitive quantum needs, 2023.

\bibitem{ayanzadeh2023enigma}
Ramin Ayanzadeh, Ahmad Mousavi, Narges Alavisamani, and Moinuddin Qureshi.
\newblock Enigma: Privacy-preserving execution of qaoa on untrusted quantum computers, 2023.

\bibitem{pulse}
Theodoros Trochatos, Chuanqi Xu, Sanjay Deshpande, Yao Lu, Yongshan Ding, and Jakub Szefer.
\newblock A quantum computer trusted execution environment.
\newblock {\em IEEE Computer Architecture Letters}, 22(2):177--180, 2023.

\bibitem{bergholm2018pennylane}
Ville Bergholm et~al.
\newblock Pennylane: Automatic differentiation of hybrid quantum-classical computations.
\newblock {\em arXiv preprint arXiv:1811.04968}, 2018.

\bibitem{paszke2019pytorch}
Adam Paszke, Sam Gross, Francisco Massa, Adam Lerer, James Bradbury, Gregory Chanan, Trevor Killeen, Zeming Lin, Natalia Gimelshein, Luca Antiga, Alban Desmaison, Andreas Köpf, Edward Yang, Zach DeVito, Martin Raison, Alykhan Tejani, Sasank Chilamkurthy, Benoit Steiner, Lu~Fang, Junjie Bai, and Soumith Chintala.
\newblock Pytorch: An imperative style, high-performance deep learning library, 2019.

\bibitem{tensorflow2015-whitepaper}
Mart\'{i}n Abadi, Ashish Agarwal, Paul Barham, Eugene Brevdo, Zhifeng Chen, Craig Citro, Greg~S. Corrado, Andy Davis, Jeffrey Dean, Matthieu Devin, Sanjay Ghemawat, Ian Goodfellow, Andrew Harp, Geoffrey Irving, Michael Isard, Yangqing Jia, Rafal Jozefowicz, Lukasz Kaiser, Manjunath Kudlur, Josh Levenberg, Dandelion Man\'{e}, Rajat Monga, Sherry Moore, Derek Murray, Chris Olah, Mike Schuster, Jonathon Shlens, Benoit Steiner, Ilya Sutskever, Kunal Talwar, Paul Tucker, Vincent Vanhoucke, Vijay Vasudevan, Fernanda Vi\'{e}gas, Oriol Vinyals, Pete Warden, Martin Wattenberg, Martin Wicke, Yuan Yu, and Xiaoqiang Zheng.
\newblock {TensorFlow}: Large-scale machine learning on heterogeneous systems, 2015.
\newblock Software available from tensorflow.org.

\bibitem{heusel2018gans}
Martin Heusel, Hubert Ramsauer, Thomas Unterthiner, Bernhard Nessler, and Sepp Hochreiter.
\newblock Gans trained by a two time-scale update rule converge to a local nash equilibrium, 2018.

\bibitem{DBLP:journals/corr/abs-1912-03817}
Lucas Bourtoule, Varun Chandrasekaran, Christopher~A. Choquette{-}Choo, Hengrui Jia, Adelin Travers, Baiwu Zhang, David Lie, and Nicolas Papernot.
\newblock Machine unlearning.
\newblock {\em CoRR}, abs/1912.03817, 2019.

\bibitem{ghost}
Xiaowan Dong, Zhuojia Shen, John Criswell, Alan Cox, and Sandhya Dwarkadas.
\newblock Spectres, virtual ghosts, and hardware support.
\newblock In {\em Proceedings of the 7th International Workshop on Hardware and Architectural Support for Security and Privacy}, HASP '18, New York, NY, USA, 2018. Association for Computing Machinery.

\bibitem{robbiati2023realtime}
Matteo Robbiati, Alejandro Sopena, Andrea Papaluca, and Stefano Carrazza.
\newblock Real-time error mitigation for variational optimization on quantum hardware, 2023.

\bibitem{Sack_2024}
Stefan~H. Sack and Daniel~J. Egger.
\newblock Large-scale quantum approximate optimization on nonplanar graphs with machine learning noise mitigation.
\newblock {\em Physical Review Research}, 6(1), March 2024.

\end{thebibliography}

\end{document}